\documentclass[12pt]{article}

\usepackage{scicite}
\usepackage{graphicx}
\usepackage{hyperref}
\usepackage{times}
\usepackage{xcolor}

\newenvironment{sciabstract}{%
\begin{quote} \bf}
{\end{quote}}

\newcounter{lastnote}

\title{The gate-tunable Josephson diode}

\author
{G.P. Mazur$^{1\ast\dagger}$, N. van Loo$^{1\dagger}$, D. van Driel$^{1}$,
J.-Y. Wang$^{1}$,\\
G.Badawy$^{2}$, S. Gazibegovic$^{2}$, E.P.A.M Bakkers$^{2}$, L.P. Kouwenhoven$^{1}$\\
\normalsize{$^{1}$ QuTech and Kavli Institute of Nanoscience,}  \\
\normalsize{Delft University of Technology, 2600 GA Delft, The Netherlands}\\
\normalsize{$^{2}$ Department of Applied Physics, Eindhoven University of Technology,} \\ 
\normalsize{5600 MB Eindhoven, The Netherlands}\\
\normalsize{2600 AD, Delft, The Netherlands}\\
\normalsize{$^\ast$To whom correspondence should be addressed. E-mail:  mazur.grzesiek@gmail.com}\\
\normalsize{$^\dagger$ These authors contributed equally to this work.}
}

\date{}

\begin{document} 


\baselineskip24pt


\maketitle

\newpage

\begin{sciabstract}
Superconducting diodes are a recently-discovered quantum analogueue of classical diodes. The superconducting diode effect relies on the breaking of both time-reversal and inversion symmetry. As a result, the critical current of a superconductor can become dependent on the direction of the applied current. The combination of these ingredients naturally occurs in proximitized semiconductors under a magnetic field, which is also predicted to give rise to exotic physics such as topological superconductivity. In this work, we use InSb nanowires proximitized by Al to investigate the superconducting diode effect. Through shadow-wall lithography, we create short Josephson junctions with gate control of both the semiconducting weak link as well as the proximitized leads. When the magnetic field is applied perpendicular to the nanowire axis, the superconducting diode effect depends on the out-of-plane angle. In particular, it is strongest along a specific angle, which we interpret as the direction of the spin-orbit field in the proximitized leads. Moreover, the electrostatic gates can be used to drastically alter this effect and even completely suppress it. Finally, we also observe a significant gate-tunable diode effect when the magnetic field is applied parallel to the nanowire axis. Due to the considerable degree of control via electrostatic gating, the semiconductor-superconductor hybrid Josephson diode emerges as a promising element for innovative superconducting circuits and computation devices.
\end{sciabstract}
\newpage
\section*{Introduction}
Semiconducting diodes are rectifiers that allow current to flow in only one direction. As such, they are ubiquitously used in conventional electronics.  Recently, superconducting analogueues of diodes have been realized in V/Nb/Ta superlattices, where the value of critical current depends on the polarity of the applied current~\cite{Ando_2020:N}. Consequently, it has been named the superconducting diode effect (SDE). The simultaneous breaking of both inversion symmetry and time-reversal symmetry has been identified as the critical ingredient for achieving the SDE~\cite{Edelstein_1996:JPCM,Wakatsuki_2017:SA}. Time-reversal symmetry is conventionally broken either through the application of a magnetic field or by using magnetic materials~\cite{Jeon_2022:NM}. Similarly, inversion symmetry is broken in systems with spin-orbit interaction (SOI), either through intrinsic structural asymmetry or via the application of electric fields. Interestingly, a correction to the critical current of superconducting films in the presence of an electric field was already proposed in the context of superconductors with an intrinsic polar axis~\cite{Edelstein_1996:JPCM}. It has the form of $\alpha(\textbf{c} \times \textbf{B}) \cdot \textbf{J}$ where $\alpha$ is the Rashba spin-orbit constant, the unit vector $\textbf{c}$ points along the electric field, $\textbf{B}$ is an external magnetic field, and $\textbf{J}$ is the supercurrent density. Such a correction can be obtained purely from the Ginzburg-Landau theory when both inversion symmetry and time-reversal symmetry are broken. 

In order to give more insight into the microscopic mechanism behind non-reciprocal critical currents, models that go beyond the Ginzburg-Landau theory have been proposed~\cite{Legg_2022:arXiv,Yuan_2022:PNAS}. These models suggest that a finite Cooper pair momentum is the underlying physical mechanism. In this picture, the energy of left and right moving carriers develops a finite Doppler shift $\pm$ \textbf{qv$_{\rm F}$} due to the finite momentum \textbf{q} acquired by the Cooper pairs\cite{Yuan_2022:PNAS}. On the other hand, theoretical studies on Josephson Junctions (JJs) based on semiconducting nanowires predicted direction-dependent critical currents in the presence of SOI and time-reversal symmetry breaking~\cite{Yokoyama_2013:JPSJ,Yokoyama_2014:PRB}. 
\newpage
\noindent{In this case, non-reciprocity of switching currents is caused by the interaction between multiple Andreev levels in the junction. In addition, the Meissner effect has been proposed~\cite{Davydova_2022:SA} to give rise to non-reciprocal critical current. Finally, a small out-of-plane magnetic field also leads to an SDE in type-II elemental superconductors~\cite{Yasen_2022:arXiv}.}

Most of the work to date has observed the SDE in metallic systems or van der Waals materials with a high electron density~\cite{Wu_2022:N}, which implies that such devices cannot be tuned electrostatically. In this context, proximitized semiconductors are a convenient platform on which various parameters can be tuned with electrostatic gates. This includes the transparency~\cite{Heedt:2021_NC} and the number of active modes in the junction, as well as the density in the proximitized region~\cite{vanLoo_2022:arXiv}. Note that in proximitized semiconductors, superconducting correlations are carried by electron-hole pairs, as opposed to a Cooper pair condensate in regular metals. These pairs form Andreev bound states which can obtain a finite momentum due to the interplay between spin-orbit interaction and a Zeeman field~\cite{Li_2018:NM,Hart_2017:NP,Pal_2022:NP}. The finite-momentum ABSs can be considered as the proximitized analogue of finite-momentum Cooper pairs. As a consequence, it is possible for an SDE to exist in a semiconductor in proximity to a regular superconductor, like Al. Indeed, recent works explored the SDE in weakly-proximitized InAs~\cite{shabani_2016:PRB}, either in the form of Josephson junctions~\cite{Baumgartner_2022:NN,Gupta_2022:arXiv} or as a metallic wire defined on the InAs stack~\cite{Sundaresh_2022:arXiv}. Planar Josephson junctions based on InSb nanoflags are also reported to yield an SDE~\cite{Turini_2022:arXiv}. Most works do not report a strong effect of electrostatic gating on the SDE~\cite{Baumgartner_2022:NN,Turini_2022:arXiv,Sundaresh_2022:arXiv,Gupta_2022:arXiv}. As such, superconducting diodes would have limited use as circuit elements of superconducting computation devices or quantum computers. Instead, a greater degree of control is desired~\cite{Leroux_2022:arxiv,Splithoff_2022:PRApp}, similar to the electrical tunability of state-of-the-art quantum electronics based on semiconductors~\cite{Zwerver_2022:NE}. In this article, we demonstrate the presence and control of the SDE in semiconductor-superconductor hybrid Josephson junctions by means of electrostatic gating. Separate gate control of the electron density underneath the proximitized leads as well as in semiconducting weak link enables us to isolate their contribution to the observed SDE. \\
\noindent{Furthermore, we show that the efficiency of the SDE scales with the switching current in the JJ. In addition, we demonstrate that the SDE can also occur when a magnetic field is applied parallel to the nanowire axis and hence parallel to the current flow. Dual gating of the semiconducting weak link and the proximitized leads allows the InSb/Al JJ to act as a Josephson field-effect transistor as well as a superconducting diode. As such, it emerges as a promising platform for the development of superconducting circuits.}

 \begin{figure}[h!]
    \centering
        \includegraphics[width=0.66\textwidth]{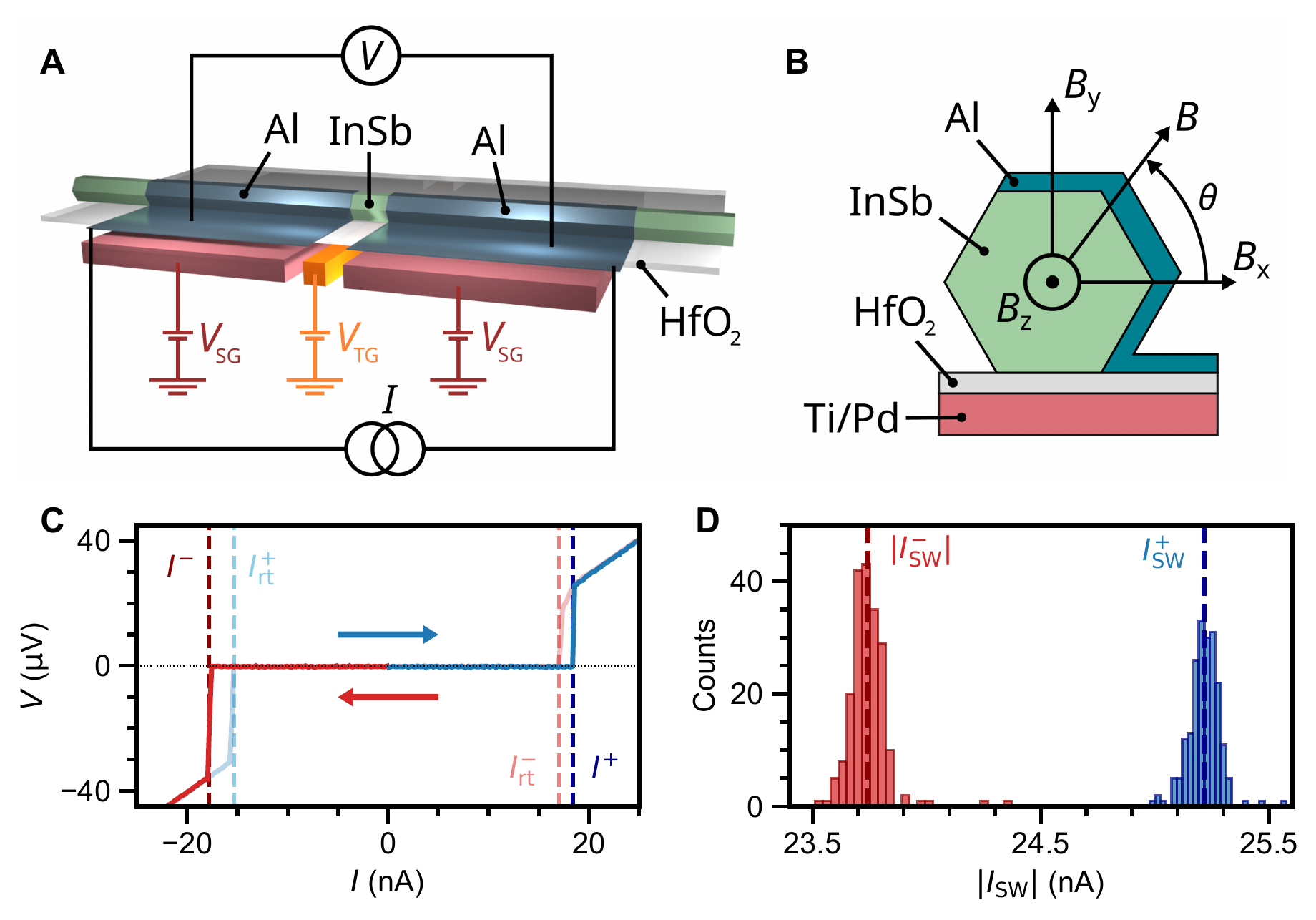}
	\caption{\textbf{Illustration of the device and measurement technique.} (\textbf{A}) Schematic of the measurement circuit, depicting a Josephson Junction in the current-bias configuration. The InSb nanowire (green) is covered by an Al shell (blue), which forms a connection to the film on the substrate. The electronic density underneath the shell is controlled via the super gates (red), and the junction transparency is tuned with the tunnel gate (orange). (\textbf{B}) Illustration of the hybrid cross-section. The magnetic field angle $\theta$ is defined as the angle in the plane perpendicular to the nanowire axis, starting at $\theta = 0^\circ$ when the applied magnetic field is along $B_{\mathrm {x}}$. (\textbf{C}) Example of $I$-$V$ curves showing non-reciprocal behavior. Blue and red curved display forward and reverse current bias, respectively. Positive and negative switching current ($I^+$,$I^-$) and retrapping current ($I_{\mathrm{rt}}^+$,$I_{\mathrm{rt}}^-$) are marked with dashed lines. (\textbf{D}) Example of switching current histograms, each consisting of switching events ($I^+$,$I^-$) from 200 $I$-$V$ curves. The average value of the distributions is labelled $|I_{\rm SW}^{\mathrm {-}}|$ and $I_{\rm SW}^{\mathrm {+}}$ for the negative and positive switching currents, respectively.} 
    \label{fig:Fig_1}
\end{figure}

\section*{Methods}
We study a Josephson junction made from an InSb nanowire coupled to an Al shell. Fig.~\ref{fig:Fig_1}A shows a schematic of such a device together with the measurement circuit. The InSb nanowire is placed on a HfO$_{2}$ dielectric, which separates the device from a set of local bottom gates. The device features two distinct gates: A voltage on the tunnel gate $V_{\rm TG}$ is used to control the semiconducting weak link, where it affects both the number of active modes as well as their transparency. On the other hand, a voltage on the super gates $V_{\rm SG}$ controls the electron density in the proximitized leads (that is, the InSb segments underneath the Al). In a recent work~\cite{vanLoo_2022:arXiv} we have shown that this gate can also be used to adjust the semiconductor-superconductor coupling, resulting in gate-tunable properties such as the induced gap and $g$-factor. A current-bias $I$ is applied through the junction, and the resulting voltage drop $V$ is measured. The sample is fabricated by means of our shadow-wall lithography technique\cite{Heedt:2021_NC,Borsoi:2021_AFM}. This fabrication method allows us to create semiconducting weak links as short as 50\,nm~\cite{Levajac_2022:Arxiv} such that the device is expected to be in the short junction limit. Since the length of the semiconducting weak link is shorter than the coherence length, transport is dominated tunnel processes via Andreev bound states. In short junctions, transport is also sensitive to properties of the proximitized leads~\cite{Beenakker_1991:PRL,Beenakker_1991:PRL_2}. Fig.~\ref{fig:Fig_1}B illustrates the cross-section of the hybrid together with the coordinate system used throughout this work.

To investigate the SDE, we apply a current bias to the sample and look for a difference in the detected critical current between the forward-bias and reverse-bias measurements. In Fig.~\ref{fig:Fig_1}C, we show an example $I$-$V$ curve under conditions where an SDE should be observed (i.e. a finite magnetic field perpendicular to the nanowire axis). As $I$ is swept from negative to positive (blue curve), the measured voltage first drops to zero at the retrapping current $I_{\mathrm{rt}}^+$. At the positive switching event $I^+$, the measured voltage jumps again to a finite value. Similarly, the negative retrapping and switching events $I_{\mathrm{rt}}^-$ and $I^-$ can be obtained by reversing the current bias polarity (red curve). In this case, a small difference between $I^+ = 18.4\,$nA and $|I^-| = 17.7\,$nA is observed, evidencing the SDE. As the switching current in Josephson junctions is stochastic~\cite{Zgirski_2019:PRAp,Coskun_2012:PRL}, we employ a fast-switching detection method~\cite{Bouman_2020:PRB,Sundaresh_2022:arXiv} in order to accurately resolve its value (see Section 1.3 of the supplemental information~\cite{Supplement}). An example of a switching current distribution obtained with this method is shown in Fig.~\ref{fig:Fig_1}D, where histograms for 200 switching events ($I^+$,$I^-$) are plotted. From these distributions, the average switching currents for the forward-bias $I_{\rm SW}^{\mathrm {+}}$ and reverse-bias $|I_{\rm SW}^{\mathrm {-}}|$ are estimated. These values of the switching current are then used for calculating the SDE efficiency $\eta = (I_{\rm SW}^{\mathrm {+}} - |I_{\rm SW}^{\mathrm {-}}|) / (I_{\rm SW}^{\mathrm {+}} + |I_{\rm SW}^{\mathrm {-}}|)$.

\begin{figure}[h!]
    \centering
        \includegraphics[width=\textwidth]{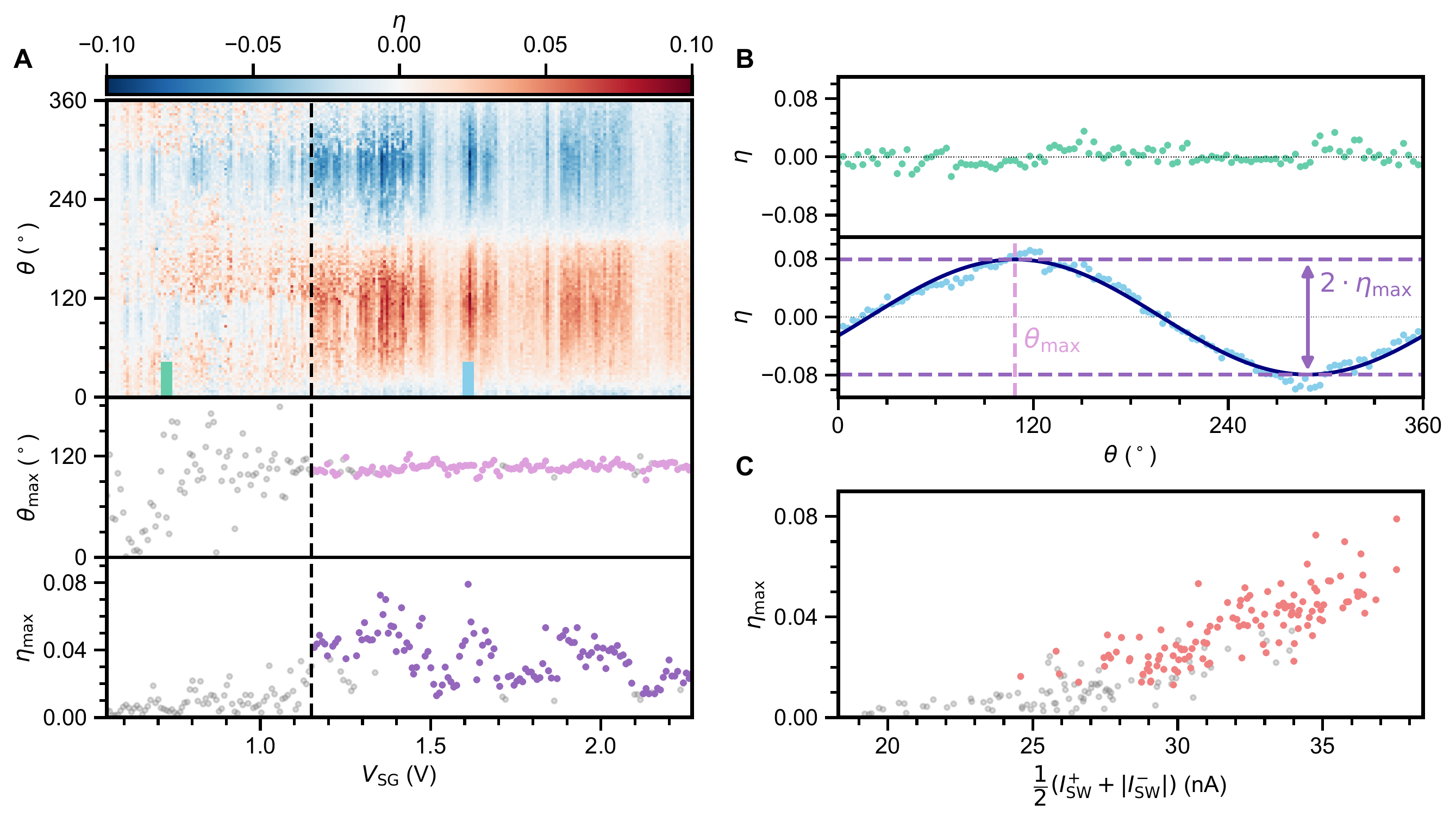}
	\caption{\textbf{Dependence of superconducting diode effect on super gate voltage.} (\textbf{A})  Top: $\eta$ as a function of super gate voltage $V_{\rm SG}$ and $\theta$, taken with $V_{\rm TG} = 3.61\,$V and $B = 12\,$mT. Middle: Estimation of $\theta_{\mathrm max}$. Bottom: $\eta_{\mathrm max}$ as a function of $V_{\rm SG}$. The dashed black line indicates that the diode effect is suppressed below $V_{\rm SG} < 1.15\,$V. (\textbf{B}) Top: lack of an SDE below $V_{\rm SG} < 1.15\,$V. Bottom: presence of an SDE above $V_{\rm SG} > 1.15\,$V. Linecuts taken from (\textbf{A}) top, at locations indicated by the colored bars. In the bottom panel, the dashed pink line specifies $\theta_{\mathrm max}$, whereas $\eta_{\mathrm max}$ is depicted by the dashed purple lines. (\textbf{C}) $\eta_{\mathrm max}$ as a function of average switching current, taken along $\theta = 105^\circ$. In all panels, grey data points correspond to a poor sinusoidal fit ($R^2 < 0.85$) of $\eta$ as a function of $\theta$.} 
    \label{fig:Fig_2}
\end{figure}

\section*{Results}
We start the investigation of the SDE in our system by rotating the magnetic field ($B = 12$\,mT) in the plane perpendicular to the nanowire axis, so that the direction of the current remains perpendicular to the applied field. We set the tunnel gate voltage to a high $V_{\rm TG}=3.61\,$V, which ensures that the junction has around $5$-$10$ active modes with a high transparency (see Fig.\,S14). Consequentially, this maximizes the switching current in the semiconducting weak link. The dependence of the diode efficiency on the magnetic field angle and super gate voltage is shown in Fig.~\ref{fig:Fig_2}A. In the top panel, we identify two distinct behaviors of the SDE. Above $V_{\rm SG} > 1.15\,$V we observe a finite diode efficiency which exhibits a sinusoidal dependence on the angle of the magnetic field. A line cut in this regime taken at $V_{\rm SG} = 1.61\,$V is shown in the bottom panel of Fig.~\ref{fig:Fig_2}B. A sinusoidal fit of $\eta(\theta)$ allows us to extract the angle with a maximum efficiency $\theta_{\rm max}$ as well as the efficiency $\eta_{\rm max}$ at that angle (see supplemental information~\cite{Supplement}). From the fit for this particular line cut, we estimate $\theta_{\rm max}=105^\circ$ and $\eta_{\rm max} = 0.08$. In contrast, the SDE is diminished and the field-angle dependence is almost absent for $V_{\rm SG}$ below $V_{\rm SG} < 1.15\,$V. This is emphasized in the top panel of Fig.~\ref{fig:Fig_2}B, which shows a line cut taken at $V_{\rm SG} = 0.72\,$V. We execute the fitting procedure for all values of $V_{\rm SG}$ in order to determine $\theta_{\rm max}$ and $\eta_{\rm max}$ as we vary the gate voltage. 
\\
\noindent{These are presented in the middle and bottom panels of Fig.~\ref{fig:Fig_2}A, respectively. We see that the estimated $\theta_{\rm max}$ angle remains roughly constant above $V_{\rm SG} > 1.15\,$V at $\theta_{\rm max} \approx 105^\circ$. We interpret this angle to be the direction of the spin-orbit field in the proximitized leads $\theta_{\rm max} = \theta_{\rm SO}$, which we elaborate on in the discussion section. In contrast, $\eta_{\rm max}$ is modulated between $\sim 1\,$\% up to $\sim 8\,$\% and does not simply increase with $V_{\rm SG}$. We note that in both panels, the grey data points correspond to a poor sinusoidal fit of $\eta(\theta)$ with a R-squared value of $R^2 \leq 0.85$.}\\
\newpage
\noindent{We refer to Fig.\,S4 in the supplemental information~\cite{Supplement} for the underlying switching current maps and the estimation of the fitting error. As $V_{\rm SG}$ also modulates the magnitude of the switching current, we plot the maximum diode efficiency versus the average switching current in Fig.~\ref{fig:Fig_2}C. Interestingly, an increase in $\eta_{\rm max}$ appears to correlate to higher switching currents. This observation naturally raises the question whether the observed SDE can be attributed to the contribution from the proximitized leads of the device or to the increasing transparency of the junction due to capacitive coupling between the gates.}

\begin{figure}[h!]
    \centering
        \includegraphics[width=\textwidth]{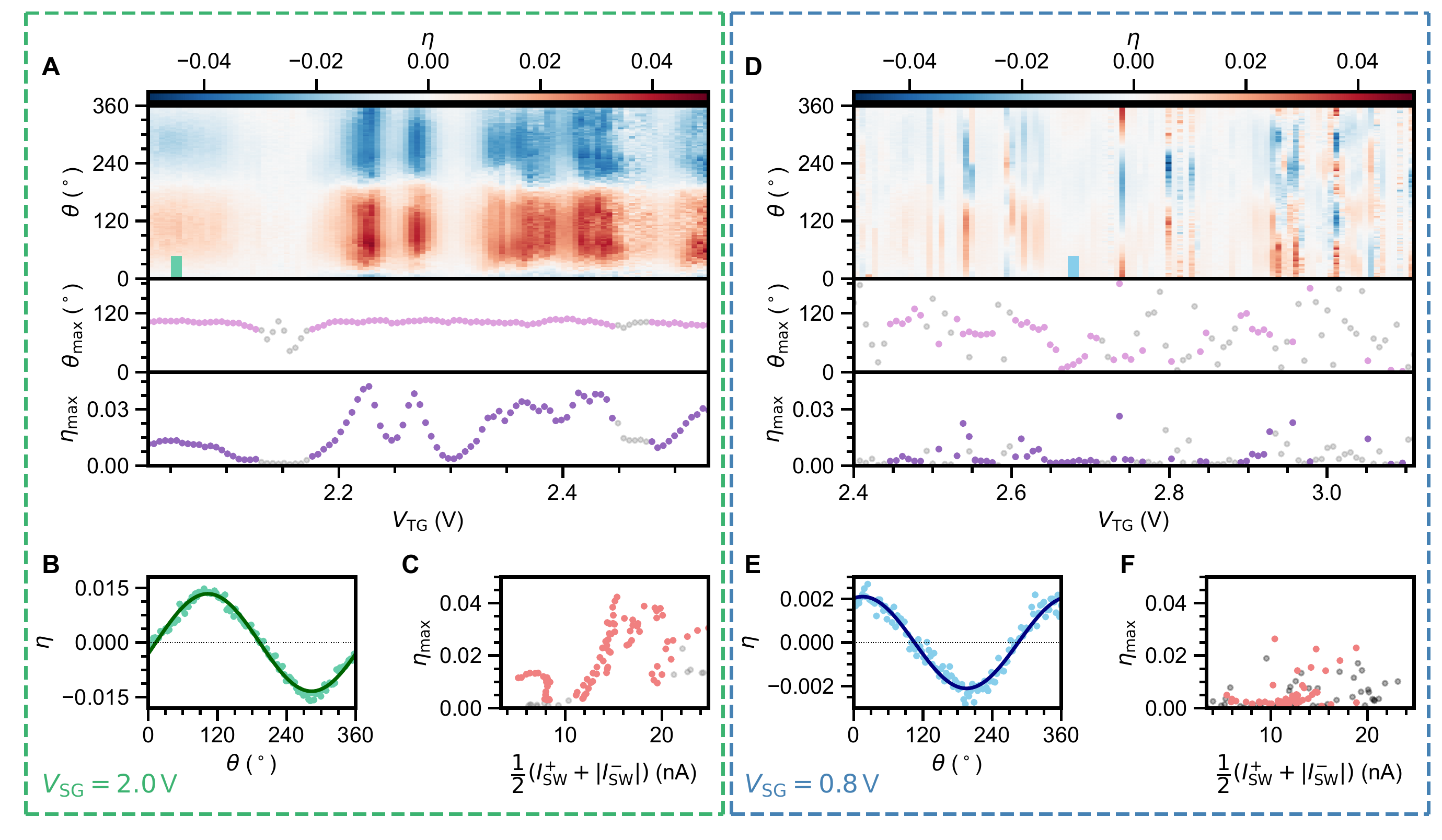}
	\caption{\textbf{Dependence of superconducting diode effect on tunnel gate voltage.} (\textbf{A-C}) Dependence on $V_{\rm TG}$ at a high super gate voltage $V_{\rm SG} = 2.0\,$V. (\textbf{A}) Top: $\eta$ as a function of $V_{\rm TG}$ and $\theta$. Middle: Estimation of $\theta_{\mathrm max}$. Bottom: $\eta_{\mathrm max}$ as a function of $V_{\rm TG}$. (\textbf{B}) Example of $\eta$ taken at the location indicated by the colored bar in panel (\textbf{A}). (\textbf{C}) $\eta_{\mathrm max}$ as a function of average switching current, taken along $\theta = 105^\circ$. (\textbf{D-F}) Dependence on $V_{\rm TG}$ at a low super gate voltage $V_{\rm SG} = 0.8\,$V. (\textbf{D}) Top: $\eta$ as a function of $V_{\rm TG}$ and $\theta$. Middle: Estimation of $\theta_{\mathrm max}$. Bottom: $\eta_{\mathrm max}$ as a function of $V_{\rm TG}$.  (\textbf{E}) Example of $\eta$ taken at the location indicated by the colored bar in panel (\textbf{D}). (\textbf{F}) $\eta_{\mathrm max}$ as a function of average switching current, taken along $\theta = 105^\circ$. In all panels, grey data points correspond to a poor sinusoidal fit ($R^2 < 0.85$) of $\eta$ as a function of $\theta$. Data taken at $B = 12\,$mT.} 
    \label{fig:Fig_3}
\end{figure}

To answer this question, we note that the transparency of the junction can be adjusted directly through the use of the tunnel gate. We proceed by fixing the super gate voltage to $V_{\rm SG} = 2$\,V, above the previously-identified threshold of $V_{\rm SG} = 1.15$\,V. Next, the tunnel gate voltage $V_{\rm TG}$ is varied such that switching current is being modulated from $|I_{\rm SW}|=5\,$nA up to $|I_{\rm SW}|=25\,$nA. We again vary the angle $\theta$ of the perpendicular magnetic field with an amplitude of $B=12\,$mT, while measuring the diode efficiency. The results of this experiment are presented in the left panel of Fig.~\ref{fig:Fig_3}, where in panel A we show the evolution of the SDE. The effect is present for almost all gate values, while $\eta$ is modulated with $V_{\rm TG}$. Interestingly, the extracted $\theta_{\rm max}$ remains constant as a function of the junction transparency around $\theta_{\rm max}=105^{\circ}$. In particular, $\eta$ remains sinusoidal in $\theta$ (see, for example, Fig.~\ref{fig:Fig_3}B), and adjustments in $V_{\rm TG}$ only affect the amplitude. The scaling of $\eta_{\rm max}$ with $I_{SW}$ is no longer monotonic, while the highest efficiency is still observed for high switching currents in the range of $|I_{\rm SW}|=15$-$20\,$nA (Fig.~\ref{fig:Fig_3} C). Moreover, we are able to measure an appreciable SDE even at very low switching currents on the order of $|I_{\rm SW}|=5\,$nA. This confirms that the observed transition in Fig.~\ref{fig:Fig_2} cannot be attributed to a reduced switching current as a result of, for example, capacitive coupling between the tunnel gate and the super gate. The situation is drastically different for $V_{\rm SG}= 0.8\,$V, which we present in the right panel of Fig.~\ref{fig:Fig_3}. Here, we adjust the tunnel gate voltage range such that it covers a similar range of switching currents, between $|I_{\rm SW}|=5\,$nA up to $|I_{\rm SW}|=25\,$nA. In Fig.~\ref{fig:Fig_3}D, we see that the efficiency of the SDE is generally weaker. It is often non-sinusoidal in $\theta$ and $\theta_{\rm max}$ varies strongly with $V_{\rm TG}$. An example is shown in Fig.~\ref{fig:Fig_3}E, where the angle with a maximum SDE is close to $\theta_{\rm max} = 0^\circ$. The observed $\eta_{\rm max}$ is mainly below $1\,$\%, even for the highest switching current values shown in Fig.~\ref{fig:Fig_3}F.

\begin{figure}[t]
    \centering
        \includegraphics[width=0.66\textwidth]{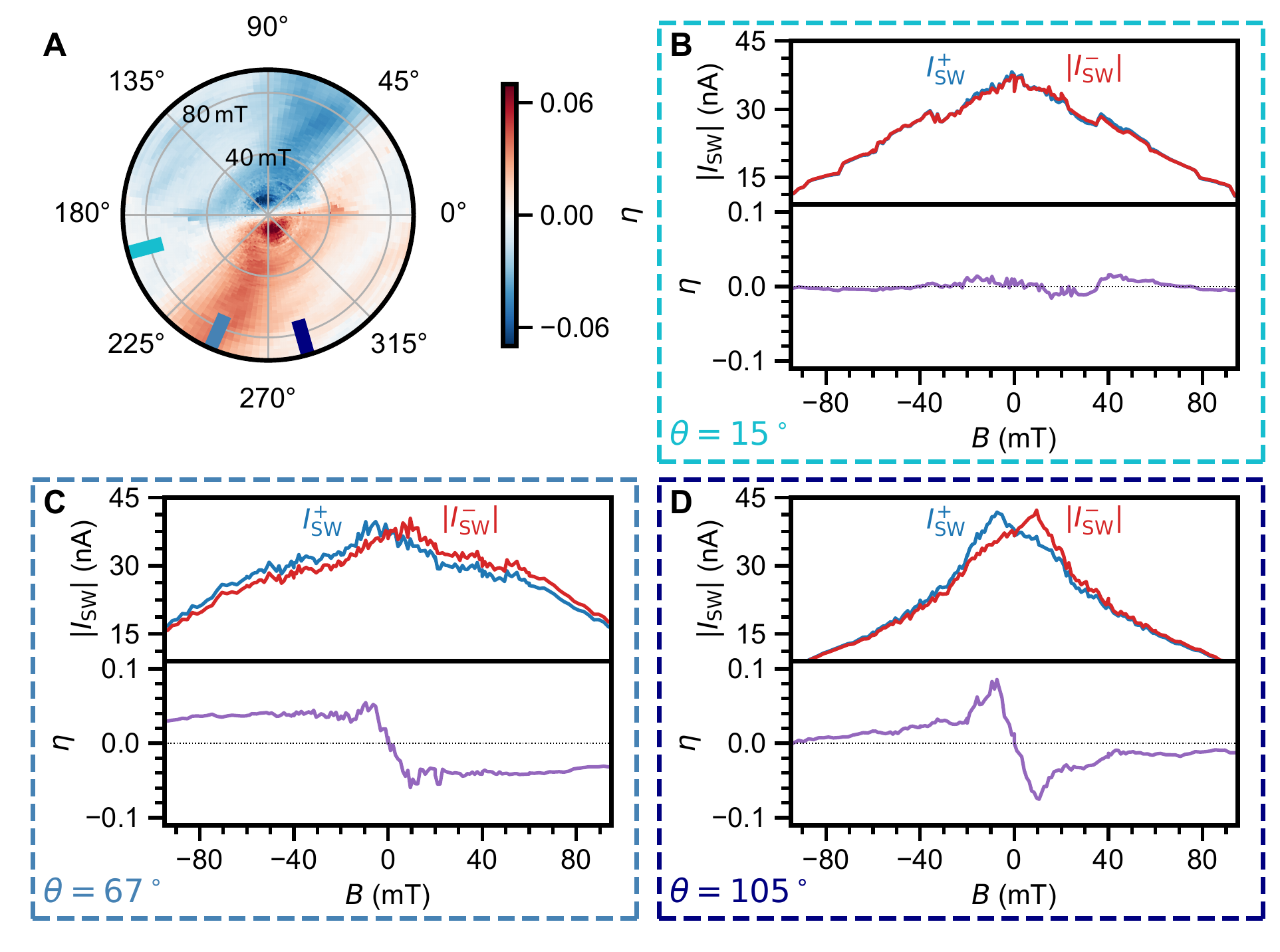}
	\caption{\textbf{Evolution of the superconducting diode effect as a function of magnetic field perpendicular to the nanowire axis.} (\textbf{A}) Polar plot of $\eta$ as a function of $\theta$ (polar axis) and field magnitude $B$ (radial axis), taken at $V_{\rm SG} = 2.21\,$V and $V_{\rm TG} = 3.60\,$V. (\textbf{B-D}) $I_{\rm SW}^{\mathrm {+}}$, $|I_{\rm SW}^{\mathrm {-}}|$ and $\eta$ as a function of $B$ taken at various angles. Perpendicular to the maximum-efficiency angle ($\theta = 15^\circ$, \textbf{B}), no SDE is observed. At $\theta = 67^\circ$ (\textbf{C}), the SDE persists in a high magnetic field. Along the maximum-efficiency angle ($\theta = 105^\circ$, \textbf{D}), the SDE increases linearly around $B = 0\,$T and reaches a maximum at $B = 10\,$mT.} 
    \label{fig:Fig_4}
\end{figure}

\begin{figure}[h!]
    \centering
        \includegraphics[width=\textwidth]{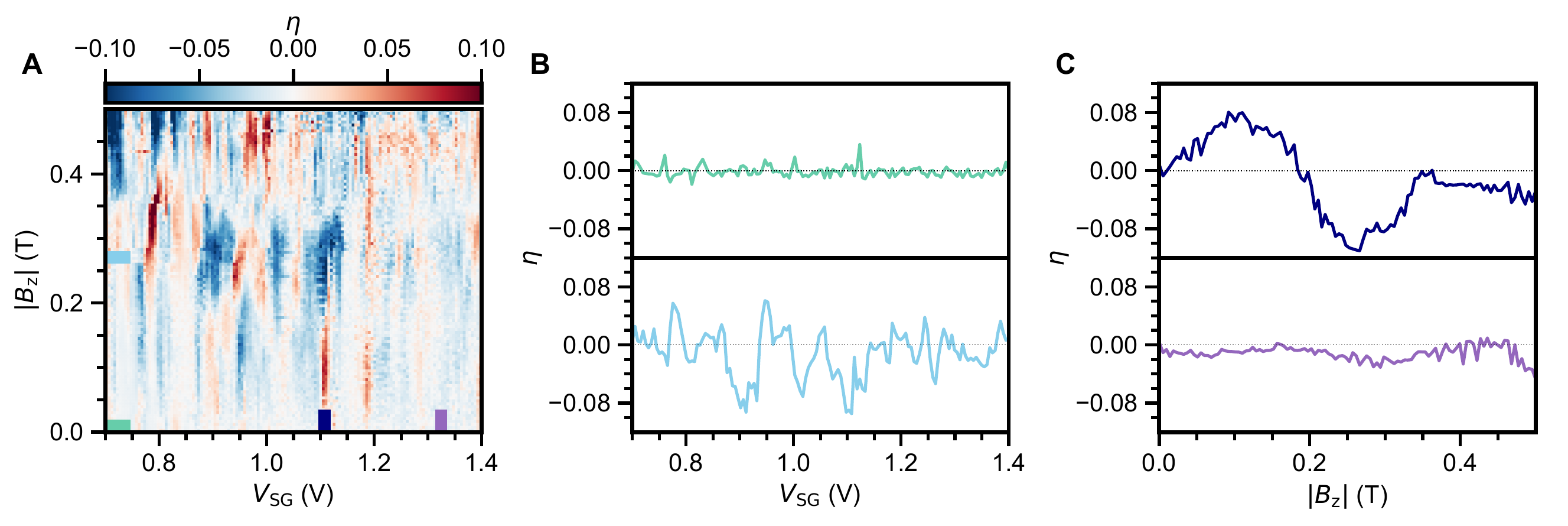}
	\caption{\textbf{Evolution of the superconducting diode effect as a function of magnetic field parallel to the nanowire axis.} (\textbf{A}) $\eta$ as a function of $V_{\rm SG}$ and $B_{\rm z}$. (\textbf{B}) Examples of $\eta$ as a function of $V_{\rm SG}$ taken at locations indicated by the colored bars in panel (\textbf{A}). At finite field (bottom), the SDE frequently changes sign and the efficiency is strongly modulated by the tunnel gate voltage. (\textbf{C}) Examples of $\eta$ as a function of $B_{\rm z}$ taken at locations indicated by the colored bars in panel (\textbf{A}). Depending on $V_{\rm SG}$, the magnetic field can both enhance the diode efficiency and cause a sign inversion.} 
    \label{fig:Fig_5}
\end{figure}

Next, we turn our attention to the dependence of the SDE on the magnitude of the magnetic field. The results are shown in Fig.~\ref{fig:Fig_4}A, where the polar axis is the field angle $\theta$ and the radial axis represents the field magnitude $B$. Line cuts at various angles are shown in Fig.~\ref{fig:Fig_4}B-D.
\\[2ex]
\noindent{In the majority of the experiments to date, the efficiency of the diode does not simply increase linearly with magnetic field. Instead, the linear dependence only holds for low values of $B$, after which it peaks at a particular field value before it decays to zero as can be seen in Fig.~\ref{fig:Fig_4}D, where the field is applied along previously identified $\theta_{\rm max}=105^{\circ}$. In our device, we also find an unexpected deviation from that behavior when the magnetic field is applied along $\theta=65^{\circ}$ (see Fig.~\ref{fig:Fig_4}C). Here, $\eta$ increases linearly before saturating around $B=10\,$mT (see Fig.\,S12 for an extended magnetic field range). In addition, we note that the SDE is negligible whenever the magnetic field is applied perpendicular to $\theta_{\rm max}$ at $\theta = 15^\circ$, as shown in Fig.~\ref{fig:Fig_4}B.}
\newpage
Lastly, we examine the impact of a parallel magnetic field on the SDE in our system. The results of this experiment are presented in Fig.~\ref{fig:Fig_5}.
At zero magnetic field (top panel in Fig.~\ref{fig:Fig_5}B), the SDE is predominantly absent with occasional spikes at conductance resonances. Upon increasing $B_{\rm z}$, the SDE becomes more pronounced and $\eta$ exhibits multiple changes of sign as a function of $V_{\rm SG}$, as shown in the bottom panel of Fig.~\ref{fig:Fig_5}B. Strikingly, our experimental data can once again be split into two different characteristic regimes. For the low super gate region below $V_{\rm SG} < 1.15\,$V, the SDE exhibits frequent polarity flips as well as a high efficiency. At voltages above that threshold, the SDE is still present, however, in general with smaller intensity and with fewer polarity switches. This is best seen by comparing top and bottom panel of Fig.~\ref{fig:Fig_5}C. The top panel shows a line cut around $V_{\rm SG}=1.1\,$V, which exhibits a large $\eta$  as well as a polarity switch. The bottom panel on the other hand presents a line cut along $V_{\rm SG}=1.3\,$V, where the SDE is strongly quenched.

\section*{Discussion}
One of the primary findings of our experimental work is a sinusoidal angle dependence of the SDE over a large range of positive $V_{\rm SG}$. Such dependence can be interpreted by taking into account the role of spin-orbit interaction proposed in Edelstein's model for polar superconductors~\cite{Edelstein_1996:JPCM}, which predicts a correction to the critical current in the form of $\alpha(\textbf{c} \times \textbf{B}) \cdot \textbf{J}$. From this prediction we expect a diode efficiency of $\eta \propto \cos(\theta)$ that is accompanied by a maximum diode efficiency whenever the magnetic field aligns with the direction of the spin-orbit field. The spin-orbit interaction in our devices likely originates from the electric field applied on the super gates. As the Al film covers three facets of the hexagonal nanowire and extends onto the substrate (Fig.~\ref{fig:Fig_1}B), the electric field lines are expected to bend towards the metallic half-shell~\cite{Bommer_2019:PRL}. This agrees well with the observed maximum-efficiency angle of $\theta_{\rm max}=105^{\circ}$. On a second device (device B, see supplemental information~\cite{Supplement}), the extracted $\theta_{\rm max}$ is similar at  $\theta_{\rm max}=110^{\circ}$. This is also in agreement with previous results, which studied the dependence of spin-orbit interaction on the induced superconducting gap~\cite{Bommer_2019:PRL} for devices in the same geometry. 

A related observation is the dependence of the SDE on the value of the super gate voltage $V_{\rm SG}$. As depicted in Fig.~\ref{fig:Fig_2}A, we notice a sharp onset in the SDE efficiency $\eta$ and the sinusoidal angular dependence. There is a striking similarity between this sharp onset of the SDE and the tunable semiconductor-superconductor coupling observed for InSb/Al hybrids~\cite{vanLoo_2022:arXiv}. The tunable coupling is reflected in a strong gate-tunability of the induced gap size as well as the $g$-factor. As the SOI is expected to be modulated by the coupling as well~\cite{Antipov:2018_PRX,Reeg_2018:PRB}, it may explain the difference between the low and high super gate regimes. We interpret our observations as follows. For high $V_{\rm SG}$ the system can be seen as S-S'-N-S'-S, where S is the Al shell, S' is the proximitized semiconductor, and N is the semiconducting weak link. In this regime, the hybrid inherits semiconducting properties such as a high $g$-factor and appreciable SOI.
\\
\noindent{The presence of a finite SOI in addition to a Zeeman field results in a finite momentum of the ABSs that form in the proximitized leads~\cite{Li_2018:NM,Hart_2017:NP,Pal_2022:NP}. In contrast, for low $V_{\rm SG}$ the hybrid can be seen as S-N-S where S represents the metallic (Al-like) leads with weak SOI. In the absence of SOI, the ABSs in the leads do not obtain a finite momentum - resulting in the suppression of the SDE. Still, the semiconducting weak link itself also possesses a strong SOI which can also lead to finite SDE. In this scenario, conductance resonances originating from confinement near the semiconducting weak link~\cite{Jarillo-Herrero_2006:N} yield a fluctuating spin-orbit direction~\cite{Lin_2022:arXiv}. Thus, we interpret the existence of erratic SDE in the low $V_{\rm SG}$ regime as originating from SOI in the semiconducting weak link.}

It is worth to mention that recently, it was shown that the SDE can also arise in systems without SOI~\cite{Davydova_2022:SA}. In this framework, screening currents present in the superconducting shell due to the Meissner effect lead to a spatially varying order parameter. This can equally well result in a formation of finite-momentum ABSs in proximitized material. In the case of our study, the maximum contribution from orbital effects in the Al shell is expected for a magnetic field direction perpendicular to the middle facet of the nanowire at $\theta=30^{\circ}$ (see Fig.~\ref{fig:Fig_1}B). Yet, we observe a maximum $\eta$ for a magnetic field angle $\theta_{\rm max}=105^{\circ}$, almost perpendicular to this direction. Moreover, $\eta$ measured along $\theta=15^{\circ}$ is close to zero in a magnetic field range of $\pm B=100\,$mT. Relating to those observations, we cannot identify the Meissner effect in the Al shell as the dominant mechanism explaining our results. Orbital effects might also have a direct effect on ABSs formed in the semiconductor, however it is presently unclear how this might influence the SDE.

The SDE dependence on the magnetic field magnitude (as shown in Fig.~\ref{fig:Fig_4}) is also pointing towards the presence of finite-momentum ABSs. In particular, the dependence of $\eta$ on $\theta$ evolves from $\eta \propto \sin(\theta)$ to $\eta \propto \sin(\pi\cdot\sin(\theta))$ as the magnetic field amplitude is increased (see also supplemental information Fig.\,S8~\cite{Supplement}). The angular dependence of the diode efficiency measured as a function of the magnitude of the magnetic field looks nearly identical compared to the results reported for NiTe$_{2}$~\cite{Pal_2022:NP}. \\
\noindent{There, this behavior was directly attributed to the presence of finite-momentum Cooper pairing. This observation suggests that the phenomenological theory of finite-momentum Cooper pairing can be universally applied to different material systems and may help to discriminate between various physical origins of the observed SDE. Proximitized InSb is particularly appealing in this context, as it has a fairly simple Fermi surface~\cite{lutchyn:2018_NRP} and the proximity effect has been widely studied in this material~\cite{Prada:2020_NRP,vanLoo_2022:arXiv} which should lead to a simplification of theoretical studies.}

Most of the models to date require the applied magnetic field to be parallel to the spin-orbit axis and perpendicular to the current flow~\cite{Edelstein_1996:JPCM,Yuan_2022:PNAS}, with finite-momentum Cooper pairing as the microscopic origin. However, there are also several predictions regarding an SDE with the magnetic field applied parallel to the current flow. We will discuss applicability of those models to the experiment presented in Fig.~\ref{fig:Fig_5}. Short nanowire Josephson junctions with SOI in the presence of a Zeeman field were theoretically studied by Yokoyama et al.~\cite{Yokoyama_2013:JPSJ,Yokoyama_2014:PRB}. When multiple conduction channels are formed within the semiconducting weak link, the Andreev levels interact and hybridize with each other due to interplay between disorder and SOI. Upon the application of an external magnetic field, they also become subject to a Zeeman splitting. As a result, time-reversal symmetry is broken and the energy of these levels is no longer equal with respect to sign inversion of the phase $E_{\rm N}(-\varphi) \neq E_{\rm N}(\varphi)$, where $E_{\rm N}$ is the energy of the Andreev levels and $\varphi$ is a phase difference across the junction. This inequality also leads to an SDE. Similarly, a recent model~\cite{Legg_2022:arXiv} predicted that in the presence of a finite out-of-plane component, an in-plane Zeeman field is expected to drive subband transitions which should manifest as an enhancement of the SDE efficiency as well as inversion of the polarity. It is of particular interest for the study of topological superconductivity, as this information could be used to identify regions in parameter space where topological superconductivity is predicted to emerge. 
\newpage
We suspect that the SDE in a parallel magnetic field (as shown in Fig.~\ref{fig:Fig_5}) primarily originates from interaction between ABSs in the semiconducting weak link. To understand why, we note that the data can once again be divided into two sections. In contrast to the case of a perpendicular magnetic field, the SDE efficiency is the strongest at low $V_{\rm SG}$ while it remains strongly modulated with $V_{\rm TG}$ (see Fig.\,S17). In this regime, the electron density in the hybrid segments of the nanowire is expected to be low~\cite{vanLoo_2022:arXiv,de_Moor_2018:NJP,Shen_2021:PRB}. Conversely, we keep $V_{\rm TG} =3.615\,$V which ensures a high density in the InSb weak link. The corresponding potential profile (e.g. a low $V_{\rm SG}$ in combination with a high $V_{\rm TG}$) can result in additional confinement for the Andreev levels formed in the semiconducting weak link. A change in the confinement has been shown to change the magnitude and direction of the SOI~\cite{Lin_2022:arXiv}, and also affects the scattering between different levels. Together, the interplay between these effects can modify the Andreev level spectrum in the weak link and, as a consequence, also affect the SDE. In particular, a phase shift $\varphi \approx \pi$ in the current-phase relation of the Andreev levels can generate a switch in the polarity of the SDE, as predicted by Yokoyama et al~\cite{Yokoyama_2014:PRB}. Such a phase shift can either originate from the changing confinement or Zeeman splitting of the Andreev levels in a magnetic field. On the other hand, Legg et al.~\cite{Legg_2022:arXiv} also predicted sign changes in the SDE polarity as the result of subband crossings in the hybrid. However, in this proposal a magnetic field component perpendicular to the nanowire axis is also required, which we do not apply in the current experiment. Thus, we cannot attribute our observations to subband physics. Furthermore, the addition of a small out-of-plane component does not significantly alter the picture (see supplemental information Fig.\,S15~\cite{Supplement}).

\section*{Conclusions}
In conclusion, we have demonstrated the existence and gate-tunability of the Josephson diode effect in proximitized InSb nanowires. We have identified that it has a strong dependence both on the electronic density in the leads, as well as in the semiconducting weak link.\\
\noindent{For a high density in the leads, the angle for which diode efficiency is maximized is fixed. We interpret this angle as the direction of the spin-orbit field in the proximitized leads, which is in agreement with previous work on devices with similar geometry~\cite{Bommer_2019:PRL}. Likewise, the semiconducting weak link can give rise to the SDE, albeit with much weaker efficiency. There, the maximum angle is strongly modulated by the tunnel gate voltage, which we assign to a modification of the confinement potential. Our measurements at high magnetic fields point to finite-momentum Andreev bound states as a microscopic mechanism for the observed diode effect, in accordance with recent theoretical proposals and experiments~\cite{Davydova_2022:SA,Pal_2022:NP}. Finally, we show that the superconducting diode effect is also present when the field is applied parallel to the nanowire axis.}

This work for the first time demonstrates the impact of the electronic density in the leads and semiconducting weak link on the SDE. As a consequence, gate-tunable superconducting diodes can be utilized as a building block of superconducting quantum devices. For example, proposals already exist which envision the use of these devices as on-chip gyrators and circulators~\cite{Leroux_2022:arxiv}. Moreover, the gate-tunable switching current allows the InSb/Al JJ to act as a Josephson field-effect transistor - establishing it as a highly versatile and promising circuit element for superconducting electronics. 
In addition, this system can be readily used to create superconducting quantum interference devices (SQUIDs), in order to investigate the current-phase relation of a Josephson diode. In this case, one arm of the loop is tuned to the regime with a strong SDE and well defined maximum SDE angle, while the reference arm can be tuned to a trivial regime without any SDE. As theoretically proposed, gate-tunable junctions can also be embedded in many-loop interferometers to achieve unprecedentedly high efficiencies~\cite{Souto_2022:arxiv}. Further improvements to the circuits can be made through embedding quantum dots in the junction, which may allow for achieving the SDE at zero magnetic field~\cite{Szombati_2016:NP}.
\section*{Data Availability and Code availability}
Raw data presented in this work and the data processing/plotting codes are available at \\ \url{https://doi.org/10.5281/zenodo.7351273}.
\section*{Acknowledgements}
We thank Raymond Schouten for the technical support on switching current measurements. We thank Liang Fu, Chun-Xiao Liu and Mert Bozkurt for helpful discussions. This work has been financially supported by the Dutch Organization for Scientific Research (NWO) and Microsoft Corporation Station Q.
\section*{Author Contribution}
G.P.M. and N.v.L. conceived the experiment. G.P.M. and N.v.L. fabricated and measured the devices. J.Y.W., D.v.D. assisted with sample fabrication and/or measurements. G.P.M. and N.v.L. analyzed the transport data. G.B. S.G and E.P.A.M.B. performed nanowire synthesis. G.P.M. and N.v.L. wrote the manuscript with valuable input from all authors. L.P.K. supervised the project.

\bibliography{scibib}

\bibliographystyle{Science}

\end{document}